\documentclass[aps,preprint]{revtex4}%
\usepackage{amsfonts}%
\usepackage{amsmath}%
\usepackage{amssymb}%
\usepackage{graphicx}

\def\figw{6cm}

\begin{document}
\title{Measurements of Stationary Josephson Current between High-$T_{c}$
Oxides as a Tool to Detect Charge Density Waves}
\author{Alexander M. Gabovich}
\affiliation{Institute of Physics, National Academy of Sciences of Ukraine, 46, Nauka
Ave., Kyiv 03028, Ukraine}
\author{Mai Suan Li}
\affiliation{Institute of Physics, Polish Academy of Sciences, 32/46, Al. Lotnik\'{o}w,
PL-02-668 Warsaw, Poland}
\author{Henryk Szymczak}
\affiliation{Institute of Physics, Polish Academy of Sciences, 32/46, Al. Lotnik\'{o}w,
PL-02-668 Warsaw, Poland}
\author{Alexander I. Voitenko}
\affiliation{Institute of Physics, National Academy of Sciences of Ukraine, 46, Nauka
Ave., Kyiv 03028, Ukraine}

\begin{abstract}
Stationary Josephson tunnel current $I_{c}$ between superconductors with
$d$-wave order parameter symmetry and charge-density-wave (CDW) partial
gapping was analyzed in the two-dimensional model appropriate to high-$T_{c}$
cuprates. It was shown that, in certain experimental setups, due to the
peculiar overlap of superconducting and CDW gaps in the momentum space, the
dependence of $I_{c}$ on the CDW parameters may be strongly nonmonotonic.
Hence, we suggested that $I_{c}$ measurements in the wide range of dopings can
serve as an indicator of CDW existence in the pseudogap regions of the cuprate
phase diagrams. Besides, the orientation $I_{c}$-dependences were analyzed.

\end{abstract}

\maketitle

\section{Introduction}

Since an unexpected and brilliant discovery of high-$T_{c}$ superconductivity
in cuprates in 1986 \cite{bednorz86:189},\ experts have been trying to find
the origin of superconductivity in them, but in vain. There are several
problems that are interconnected and probably cannot be solved independently.
But they are so complex that researchers are forced to consider them
separately in order to find the key concepts and express key ideas explaining
the huge totality of experimental data. General discussion and the analysis of
high-$T_{c}$-oxide superconductivity can be found in comprehensive reviews
\cite{fisk11:book,rice12:016502,yoshida12:011006,kamimura12:677,sebastian12:102501,tafuri13:21,sacuto13:022502,tranquada13:114,maiti13:3,keimer13:1,pickett13:1,hashimoto14:483,das14:151}%
. In particular, the main questions to be solved are as follows: (i) Is
superconductivity in cuprates a conventional one based on the Cooper pairing
concept? (ii) If the answer to the first question is positive, what is the
mechanism of superconductivity, i.e., what are the virtual bosons that glue
electrons in pairs? (iii) Which is the symmetry of the superconducting order
parameter? This question remains unanswered, although the majority of the
researchers in the field think believe that the problem is already resolved
(namely, $d_{x^{2}-y^{2}}$-one, see, e.g., Refs.
\cite{tsuei08:869,kirtley11:436})? (iv) What is the role of the intrinsic
disorder and non-stoichiometry in the superconducting properties
\cite{mostovoy05:224502,zhou07:076401,yang09:174505,rice12:016502,he14:608,nie14:7980,fujita14:612,achkar14:107002}%
? (v) What is the origin of the symmetry loss and, specifically, the emerging
nematicity
\cite{yang09:174505,lawler10:347,rice12:016502,nie14:7980,fujita14:612,toda14:094513}%
? (vi) What is the origin of the so-called pseudogap
\cite{vishik12:18332,rice12:016502,yoshida12:011006,wahl12:514,hashimoto14:483,morr14:382}%
? (vii) What is the role of spin- and charge- density waves (SDWs and CDWs)
both in the normal and superconducting states of cuprates? The role of various
electron spectrum instabilities competing with the Cooper pairing below the
critical temperature $T_{c}$ is a part of the more general problem: How can
certain anomalous high-$T_{c}$ oxide properties above $T_{c} $ be explained,
e.g., the linear behavior of the resistivity
\cite{varma89:1996,maksimov00:1033}? In this connection, a quite reasonable
viewpoint was expressed that if one understands the normal state of cuprates,
the superconducting state properties will be perceived
\cite{hussey08:123201,hashimoto14:483}. Here, it is also worth to mention a
possible failure \cite{varma89:1996,varma02:267}\ of the Fermi liquid concept
belonging to Landau \cite{agd}\ and the role of strong electron correlations
\cite{dagotto05:257,lee06:17,lee08:012501,izyumov08:25}.

During last decades we have been developing a phenomenological theory to
elucidate the influence of CDWs on superconductivity of high-$T_{c}$ oxides,
since the CDWs were observed in a number of those materials
\cite{gabovich97:3901,gabovich01:1,gabovich02:583,gabovich10:681070,ekino11:699,gabovich13:301}%
. We identified the CDW energy gap with the pseudogap mentioned above. Such an
identification is based, in particular, on the appearance of CDWs only below
the approximate border of the pseudogapped region\ in La$_{2-x}$Sr$_{x}
$CuO$_{4}$ \cite{croft14:224513}\ and YBa$_{2}$Cu$_{3}$O$_{7-\delta}$
\cite{bakr13:214517,hucker14:054514}. Moreover, the symmetry of the pseudogap
order parameter (isotropic) differs from that for the superconducting one
($d_{x^{2}-y^{2}}$) in Bi$_{2}$Sr$_{2}$CaCuO$_{8+\delta}$
\cite{sakai13:107001}, superconductivity in Bi$_{2}$Sr$_{2-x}$La$_{x}%
$CuO$_{6+\delta}$\ emerges with doping when the (nodal) pseudogap disappears
\cite{peng13:2459}, the pseudogap competes with the superconducting gap at
antinodes in (Bi,Pb)$_{2}$(Sr,La)$_{2}$CuO$_{6+\delta}$ \cite{he14:608}, and
the interplay of pseudogapping and superconductivity among different members
of the oxide family (Ca$_{x}$La$_{1-x}$)(Ba$_{1.75-x}$La$_{0.25+x}$)Cu$_{3}%
$O$_{y}$ is not the same for varying dopings $x$ \cite{cvitanic14:054508}. It
is worthy of note that both angle-resolved photoemission spectroscopy (ARPES)
and scanning tunnel microscopy (STM) experiments allow one to measure only
overall energy gaps whatever their microscopic origin. That is why it is
usually difficult to distinguish for sure between superconducting, SDW, and
CDW gaps even in the case when they manifest themselves separately in certain
momentum ranges each \cite{boyer07:802,hashimoto14:483}.

As for direct experiments confirming the existence of CDWs competing with
superconductivity in cuprates, CDWs have been shown to be a more important
factor in this sense than SDWs, the remnants of which survive far from the
antiferromagnetic state appropriate to zero-doped samples of superconducting
families \cite{damascelli03:473}. It is useful to shortly summarize the main
new findings in this area.

X-ray scattering experiments in YBa$_{2}$Cu$_{3}$O$_{6+x}$\ revealed the CDW
ordering at temperatures lower than those of the pseudogap formation, giant
phonon anomalies, and elastic central peak induced by nanodomain CDWs
\cite{blackburn13:054506,letacon14:52,hucker14:054514,blancocanosa14:054513}.
The CDW correlation length increases with the temperature, $T$,\ lowering.
However, the competing superconducting order parameter, which emerges below
$T_{c}$, so depresses CDWs that the true CDW long-range order does not
develop, as was shown by Raman scattering \cite{bakr13:214517}. Suppression of
CDWs by Cooper pairing was also found in x-ray measurements of La$_{2-x}%
$Sr$_{x}$CuO$_{4}$\ \cite{croft14:224513}.

The well-known CDW manifestations in Bi$_{2}$Sr$_{2-x}$La$_{x}$CuO$_{6+\delta
}$ were recently confirmed by complex X-ray, ARPES, and STM\ studies
\cite{comin14:390}. Those authors associate CDWs with pseudogapping, but argue
that the CDW wave vector connects the Fermi arc tips rather than the antinodal
Fermi surface (FS) sections, as stems from the Peierls-insulator scenario
\cite{friend79:1441,gruner94:book}. This conclusion, if being true, makes the
whole picture even more enigmatic than in the conventional density-wave
approach to pseudogaps either in the mean-field approximation or taking into
account fluctuations.

The electron-hole asymmetric CDW ordering was demonstrated by STM and resonant
elastic x-ray scattering measurements \cite{dasilvaneto14:393} for Bi$_{2}%
$Sr$_{2}$CaCuO$_{8+\delta}$ samples, with the pseudogapping in the antinodal
momentum region. As was shown in those experiments, CDWs and concomitant
periodic crystal lattice distortions, PLDs can be observed directly, whereas
their interplay with superconductivity manifestations can be seen only
indirectly, e.g., as anticorrelations between $T_{c}$ and the structural,
$T_{s}$, or CDW, $T_{\mathrm{CDW}}$, transition temperature.(There is a
viewpoint \cite{dai14:165140} that the strong interrelation between electronic
CDW modulations and PLDs \cite{friend79:1441}, inherent, e.g., to the Peierls
model of the structural phase transition \cite{gruner94:book}, does not exist,
and PLDs can emerge without electronic contributions, which seems strange in
the context of indispensable Coulomb forces.). This fact is well known, say,
for superconducting transition metal dichalcogenides \cite{jerome76:125} or
pseudoternary systems (Lu$_{1-x}$Sc$_{x}$)$_{5}$Ir$_{4}$Si$_{10}$
\cite{yang91:7681}. Therefore, it seems interesting to propose such studies of
superconducting properties, which would demonstrate manifestations of CDW
existence, although the CDW gapping is an insulating rather than a
superconducting one. In a number of publications, we suggested that certain
measurements of the stationary Josephson critical current, $I_{c}$, between
quasi-two-dimensional CDW superconductors with the $d_{x^{2}-y^{2}}$ order
parameter symmetry (inherent to cuprates) can conspicuously reveal such
dependences that would reflect CDW gapping as well or at least demonstrate
that the actual gapping symmetry differs from the pure $d_{x^{2}-y^{2}}$
one\textrm{\ }%
\cite{gabovich12:289,gabovich12:414,gabovich13:301,gabovich13:104503,gabovich14:284}%
. Below, we present further theoretical studies in this direction, which put
forward even more effective experiments.


\section{Formulation}

Following the dominating idea (see our previous publications
\cite{voitenko10:1300,voitenko10:20,gabovich12:289,gabovich12:414,gabovich13:301,gabovich13:104503,gabovich14:284}
and references therein) concerning the electron spectrum of high-$T_{c}$
oxides identified as partially gapped CDW superconductors, CDWSs, we restrict
our consideration to the two-dimensional case with the corresponding FS shown
in Fig.~1a. The superconducting $d$-wave order parameter $\Delta$ is assumed
to span the whole FS, whereas the $s$-wave mean-field dielectric (CDW) order
parameter $\Sigma$ develops only on the nested (dielectrized, d) FS sections.
There are $N=4$ or 2 of the latter (the checkerboard and unidirectional
configurations, respectively), and they are connected in pairs by the
CDW-vectors $\mathbf{Q}$'s in the momentum space. The non-nested sections
remain non-dielectrized (nd). The orientations of $\mathbf{Q}$'s are assumed
to be fixed with respect to the crystal lattice. In particular, they are
considered to be directed along the $\mathbf{k}_{x}$- and $\mathbf{k}_{y}%
$-axes in the momentum space (anti-nodal nesting)
\cite{markiewicz97:1179,gabovich10:681070,plakida10:book}. The same
orientation along $\mathbf{k}_{x}$- and $\mathbf{k}_{y}$-axes is also
appropriate to $\Delta$-lobes, so that we confine ourselves to the
$d_{x^{2}-y^{2}}$-wave symmetry of the superconducting order parameter as the
only one found in the experiments for cuprates. Hence, the profile of the
$d$-wave superconducting order parameter over the FS is written down in the
form
\begin{equation}
\bar{\Delta}(T,\theta)=\Delta(T)f_{\Delta}(\theta).\label{Delta(T,teta)}%
\end{equation}
The function $\Delta(T)$ is the $T$-dependent magnitude of the superconducting
gap, and the angular factor $f_{\Delta}(\theta)$ looks like
\begin{equation}
f_{\Delta}(\theta)=\cos2\theta.\label{fDelta(teta)}%
\end{equation}

In the case $N=4$, the experimentally measured magnitudes of the CDW order
parameter $\Sigma$ in high-$T_{c}$ oxides are identical in all four CDW
sectors, and the corresponding sector-connecting $\mathbf{Q}$ vectors are
oriented normally to each other. Therefore, we assume the CDWs to possess the
four- (the checkerboard configuration) or the two-fold (the unidirectional
configuration) symmetry
\cite{gabovich10:681070,ekino11:699,ekino11:385701,gabovich13:104503,vignolle13:39,hosur13:115116,alldredge13:104520}%
. The latter is frequently associated with the electronic nematic, smectic or
more complex ordering
\cite{vojta09:699,fradkin10:153,su11:220506,kee13:202201,dasilvaneto13:161117,wang13:063007,wang13:073039,sugai13:475701,davis13:17623,fujita14:612,nie14:7980,toda14:094513}%
).\textrm{\ }The opening angle of each CDW sector, where $\Sigma\neq0$, equals
$2\alpha$. Such a profile of $\Sigma$ over the FS can also be described in the
factorized form as
\begin{equation}
\bar{\Sigma}(T,\theta)=\Sigma(T)f_{\Sigma}(\theta),\label{Sigma(T,teta)}%
\end{equation}
where $\Sigma(T)$ is the $T$-dependent CDW order parameter, and the
angular\ factor
\begin{equation}
f_{\Sigma}(\theta)=\left\{
\begin{array}
[c]{lll}%
1 & \mathrm{for~}\left\vert \theta-k\Omega\right\vert <\alpha &
\text{\textrm{(d section),}}\\
0 & \mathrm{otherwise} & \text{\textrm{(nd section).}}%
\end{array}
\right. \label{fSigma(teta)}%
\end{equation}
Here, $k$ is an integer number, and the parameter $\Omega=\pi/2$ for $N=4$ and
$\pi$ for $N=2$.

The both gapping mechanisms (superconducting and CDW-driven) suppress each
other, because they compete for the same quasiparticle states near the FS. As
a result, a combined gap (the gap rose in the momentum space, see Fig.~1b)
\begin{equation}
\bar{D}(T,\theta)=\sqrt{\bar{\Sigma}^{2}(T,\theta)+\bar{\Delta}^{2}(T,\theta
)},\label{D(T_teta)}%
\end{equation}
arises on the FS. The actual $\Delta(T)$- and $\Sigma(T)$-values are
determined from a system of self-consistent equations. The relevant initial
parameters, besides $N$ and $\alpha$, include the constants of superconducting
and electron-hole couplings recalculated into the pure BCS (no CDWs) and CDW
(no superconductivity) limiting cases as the corresponding $\Delta_{0}$ and
$\Sigma_{0}$ order parameters at $T=0$. It should be emphasized that our model
is a simplified, generic one, because real CDWs are complex objects, which
behave differently on the crystal surfaces and in the bulk \cite{rosen13:1977}%
. Thus, it is quite natural that they are not identical for various
high-$T_{c}$ oxides \cite{cvitanic14:054508}. Nevertheless, the presented
model allows the main features of the materials concerned to be taken into
account. For brevity, we mark the CDW $d$-wave superconductor with $N$ CDW
sectors as $S_{\mathrm{CDW}N}^{d}$.

The $s$-wave BCS superconductor is described in the framework of the standard
BCS theory. Its characteristic parameter is the value of the corresponding
superconducting order parameter $\Delta_{\mathrm{BCS}}$ at $T=0 $. Also for
the sake of brevity, it will be marked below as $S_{\mathrm{BCS}}^{s}$.

In the tunnel Hamiltonian approximation, the stationary Josephson critical
current is given by the formula
\cite{kulik72:book,barone82:book,tafuri05:2573}%
\begin{equation}
I_{c}(T)=4eT%
{\displaystyle\sum\limits_{\mathbf{pq}}}
\left\vert \widetilde{T}_{\mathbf{pq}}\right\vert ^{2}%
{\displaystyle\sum\limits_{\omega_{n}}}
\mathsf{F}^{+}(\mathbf{p;}\omega_{n})\mathsf{F}^{\prime}(\mathbf{q;-}%
\omega_{n}).\label{ICGen}%
\end{equation}
Here, $\widetilde{T}_{\mathbf{pq}}$ are the tunnel Hamiltonian matrix
elements, $\mathbf{p}$ and $\mathbf{q}$ are the transferred momenta; $e>0$ is
the elementary electrical charge, and $\mathsf{F}(\mathbf{p;}\omega_{n})$ and
$\mathsf{F}^{\prime}(\mathbf{q;-}\omega_{n})$ are Gor'kov Green's functions
for superconductors to the left and to the right, respectively, from the
tunnel barrier (hereafter, all primed quantities are associated with the right
hand side electrode). The internal summation is carried out over the discrete
fermionic \textquotedblleft frequencies\textquotedblright\ $\omega_{n}=\left(
2n+1\right)  \pi T$, $n=0,\pm1,\pm2,\ldots$. Below, we consider tunnel
junctions of two types: symmetric $S_{\mathrm{CDW}N}^{d}-I-S_{\mathrm{CDW}%
N}^{d}$ between two identical CDWSs, and nonsymmetric $S_{\mathrm{CDW}N}%
^{d}-I-S_{\mathrm{BCS}}^{s}$ between a CDWS as the left electrode and an
$s$-wave BCS superconductor as the right one (here, $I$ stands for the
insulator). Expressions for the corresponding Green's functions can be found
elsewhere \cite{gabovich13:301,gabovich13:104503}. Since CDWS electrodes are
anisotropic, their orientations with respect to the junction plane will be
characterized by the angles $\gamma$ and $\gamma^{\prime}$ (the latter appears
only in the symmetric case), i.e. the deflections of the \textquotedblleft
positive\textquotedblright\ $\Delta$- and $\Delta^{\prime}$-lobes from the
normal $\mathbf{n}$ to the junction (Fig.~2). Accordingly, the angular
dependences $f_{\Delta}(\theta)$ and $f_{\Sigma}(\theta)$ of the corresponding
order parameters (see formulas (\ref{fDelta(teta)}) and (\ref{fSigma(teta)},
respectively) should be modified by changing $\theta$ to $\theta-\gamma$ or
$\theta-\gamma^{\prime}$.

An important issue while calculating the Josephson current is tunnel
directionality \cite{wolf85:book}, which should be taken into consideration in
the tunnel Hamiltonian $\widetilde{T}_{\mathbf{pq}}$. Indeed, if we calculate
$I_{c}$ between, e.g., pure BCS $d$-wave superconductors, $S_{\mathrm{BCS}%
}^{d}$, making no allowance for this factor, formula (\ref{ICGen}) would
produce an exact zero. It is so because, owing to the alternating signs of
superconducting lobes, the current contributions from the FS points described
by the angles $\theta$ and $\theta+\dfrac{\pi}{2}$ would exactly compensate
each other in this case. The same situation also takes place in the case of a
junction with $S_{\mathrm{CDW}4}^{d}$. For a junction with $S_{\mathrm{CDW}%
2}^{d}$, it is not so, but, in the framework of the general approach, we have
to introduce tunnel directionality in this case as well.

Here, we briefly consider three factors responsible for tunnel directionality
(see a more thorough discussion in Ref. \cite{gabovich14:284}). First, the
velocity component normal to the junction should be taken into account. This
circumstance is reflected by the $\cos\theta$-factor in the integrand and an
angle-independent factor that can be incorporated into the junction
normal-state resistance $R_{N}$ \cite{kashiwaya00:1641,lofwander01:53}.
Second, superconducting pairs that cross the barrier at different angles
penetrate through barriers with different effective widths
\cite{bruder95:12904} (the height of the junction barrier is assumed to be
much larger than the relevant quasiparticle energies, so that this height may
be considered constant). Since the actual $\theta$-dependences of
$\widetilde{T}_{\mathbf{pq}}$\ for realistic junctions are not known, we
simulate the barrier-associated directionality by the phenomenological
function%
\begin{equation}
w(\theta)=\exp\left[  -\left(  \dfrac{\tan\theta}{\tan\theta_{0}}\right)
^{2}\ln2\right]  ,\label{wDirect}%
\end{equation}
This means that the effective opening of relevant tunnel angles equals
$2\theta_{0}$. The barrier transparency is normalized by the maximum value
obtained for the normal tunneling with respect to the junction plane and
included into the junction resistance $R_{N}$. Hence, $w(\theta=0)=1$. The
multiplier $\ln2$ in (\ref{wDirect}) was selected to provide $w(\theta
=\theta_{0})=\frac{1}{2}$. Third, we use the model of coherent tunneling
\cite{ledvij95:3269,bruder95:12904,klemm05:801}, when the superconducting
pairs are allowed to tunnel between the points on the FSs of different
electrodes characterized by the same angle $\theta$.

As a result of the standard calculation procedure
\cite{kulik72:book,barone82:book} applied to formula (\ref{ICGen}) and in the
framework of the approximations made above, we obtain the following formula
for the stationary Josephson critical current across the tunnel junction:%
\begin{align}
I_{c}(T,\gamma,\gamma^{\prime}) &  =\frac{1}{2eR_{N}}\nonumber\\
&  \times\frac{1}{\pi}\int_{-\pi/2}^{\pi/2}\cos\theta~w(\theta)~P(T,\theta
,\gamma,\gamma^{\prime})\mathrm{d}\theta,\label{IcGen}%
\end{align}
where \cite{gorbonosov67:803,gabovich90:293}%
\begin{equation}
P(T,\theta,\gamma,\gamma^{\prime})=\bar{\Delta}\bar{\Delta}^{\prime}%
\int\limits_{\min\left\{  \bar{D},\bar{D}^{\prime}\right\}  }^{\max\left\{
\bar{D},\bar{D}^{\prime}\right\}  }\frac{\tanh\frac{x}{2T}\mathrm{d}x}%
{\sqrt{\left(  x^{2}-\bar{D}^{2}\right)  \left(  \bar{D}^{\prime2}%
-x^{2}\right)  }}.\label{P(T)}%
\end{equation}
Here, for brevity, we omitted the arguments in the dependences $\bar{\Delta
}(T,\theta-\gamma)$, $\bar{\Delta}^{\prime}(T,\theta-\gamma^{\prime})$,
$\bar{D}(T,\theta-\gamma)$, and $\bar{D}^{\prime}(T,\theta-\gamma^{\prime})$.
Integration over $\theta$ in Eq.~(\ref{IcGen}) is carried out within the
interval $-\frac{\pi}{2}\leq\theta\leq\frac{\pi}{2}$, i.e. over the
\textquotedblleft FS semicircle\textquotedblright\ turned towards the junction
plane. If any directionality and CDW gapping are excluded (so that the
integration over $\theta$ is reduced to a factor of $\pi$) and the angular
factors $f_{\Delta}$ and $f_{\Delta}^{\prime}$ remain preserved, we arrive at
the Sigrist--Rice model \cite{sigrist92:4283}.


\section{Results and their discussion}

The influence of various problem parameters on the critical stationary
Josephson curent in the symmetric, $S_{\mathrm{CDW}N}^{d}-I-S_{\mathrm{CDW}%
N}^{d}$, and nonsymmetric, $S_{\mathrm{CDW}N}^{d}-I-S_{\mathrm{BCS}}^{s}$,
junctions was analyzed in detail in works
\cite{gabovich13:104503,gabovich14:284}. Here, we attract attention to the
problem of CDW detection in high-$T_{c}$ oxides.

The number of problem parameters can be diminished by normalizing the
\textquotedblleft order parameter\textquotedblright\ quantities by one of
them. For such a normalization, we selected the parameter $\Delta_{0}$ and
introduced the dimensionless order parameters $\sigma_{0}=\Sigma_{0}%
/\Delta_{0}$ and $\delta_{\mathrm{BCS}}=\Delta_{\mathrm{BCS}}(T\rightarrow
0)/\Delta_{0}$ (for the superconducting order parameter of CDWS, $\delta
_{0}=\Delta_{0}/\Delta_{0}=1$). With regard to experimental needs, we also
introduced the reduced temperature $\tau=T/T_{c}$. Here $T_{c}$ is the actual
critical temperature of the CDWS. In the framework of our theory, it has to be
found from the system of equations for the CDWS indicated above. For the
Josephson current amplitude $I_{c}$, we introduced the dimensionless
combination $i_{c}=I_{c}eR_{N}/\Delta_{0}$.

One more preliminary remark concerns the parameter of effective tunnel
directionality $\theta_{0}$ (see formula (\ref{wDirect})). Our calculations
\cite{gabovich13:104503,gabovich14:284} showed that its choice is very
important. On the one hand, large values of this parameter correspond to thin
junctions and large values of the tunnel current, which is beneficial for the
experiment. However, in this case, the predicted phenomena become effectively
smoothed out up to their disappearance. On the other hand, narrow tunnel cones
(small $\theta_{0}$-values) provide well pronounced effects, but correspond to
thick interelectrode layers and, as a result, small tunnel currents. Hence, in
the real experiment, a reasonable compromise should be found between those two extremes.


\subsection{Electrode rotation}

While examining Fig.~2, it becomes clear that the clearest way to prove that
electrons in high-$T_{c}$ oxides undergo an additional pairing of some origin
besides the $d$-wave BCS one is to demonstrate that the gap rose differs from
that in the $S_{\mathrm{BCS}}^{d}$ superconductor. The case in question
concerns pairing symmetries, which may be different from the $d$-wave one
or/and extend over only certain FS regions. In the framework of the tunnel
technique, the most direct way to perform the search is to fix one electrode
and rotate the other one (e.g., $\gamma^{\prime}=\mathrm{const}$ and
$\gamma=\mathrm{var}$). In the case of $S_{\mathrm{BCS}}^{d}-I-S_{\mathrm{BCS}%
}^{d}$ junction, the corresponding $i_{c}(\gamma)$ dependences are known to
have a cosine profile stemming from dependence (\ref{fDelta(teta)}) for the
superconducting order parameter $\Delta$ and, since any other gapping is
absent, for the corresponding gap rose ($\bar{D}(T,\theta)=\left\vert
\Delta(T,\theta)\right\vert $). Any deviations of the gap rose from this
behavior will testify in favor of the existence of additional order
parameter(s). Certainly, averaging the current over the FS will smooth the
relevant peculiarities and making allowance for tunnel directionality will
distort them. Nevertheless, the proposed method will be sufficient to detect
the competing pairing without its ultimate identification.

In Fig.~3, the corresponding normalized $i_{c}(\gamma)$ dependences calculated
for the symmetric S$_{\mathrm{CDW}N}^{d}-I-$S$_{\mathrm{CDW}N}^{d} $ junction
and the CDW geometries $N=2$ and 4, as well as the reference $d$-wave BCS
curve, are shown. The tunnel directionality parameter $\theta_{0}=10^{\circ}$
was assumed. A more detailed analysis of $i_{c}(\gamma)$ dependences and their
relations with other problem parameters can be found in work
\cite{gabovich13:104503}. The results obtained testify that the formulated
task is feasible. An attractive feature of this technique is that, instead of
the fixed $S_{\mathrm{CDW}N}^{d}$ electrode, we may use the $S_{\mathrm{BCS}%
}^{s}$ one as well, which might be more convenient from the experimental point
of view.


\subsection{Anomalous temperature dependence of $I_{c}$}

The measurement of the temperature dependences of the critical Josephson
tunnel current $I_{c}(T)$ seems to be the most easily realizable method of
those proposed in this work. The dependence $I_{c}(T)$ in the symmetric
$S_{\mathrm{BCS}}^{s}-I-S_{\mathrm{BCS}}^{s}$ junctions has a monotonic convex
shape. Among other things, this fact is associated with the constant sign of
order parameter over the whole FS. However, in the case of symmetric
$S_{\mathrm{BCS}}^{d}-I-S_{\mathrm{BCS}}^{d}$ junctions, the situation may
change. Indeed, for junctions involving YBa$_{2}$Cu$_{3}$O$_{7-\delta}$,
nonmonotonic $I_{c}(T)$-dependences and even the change of $I_{c}$ sign, i.e.
the transformation of the 0-junction into the $\pi$-one or vice versa were
observed \cite{ilichev01:5369,testa05:134520}. Such a phenomenon was not found
for other cuprates. However, it is extremely difficult to produce Josephson
junctions made of other materials than YBa$_{2}$Cu$_{3}$O$_{7-\delta}$.
Therefore, further technological breakthrough is needed to make sure that the
non-monotonic behavior is a general phenomenon inherent to all high-$T_{c}$
oxides with $d$-wave superconducting order parameter.

It should be noted that, in the measurements concerned, the electrodes
remained fixed, so that the peculiar behavior of $I_{c}(T)$ could not result
from the change of overlapping between the superconducting lobes with
different signs. There is an explanation based on the existence of the bound
states in the junction due to the Andreev--Saint-James effect
\cite{kashiwaya00:1641,lofwander01:53}. This theory predicts that the current
$I_{c}(T)$ between $d$-wave superconductors must exhibit a singularity at
$T\rightarrow0$. Nevertheless, the latter has not been observed experimentally
until now. Probably, this effect is wiped out by the roughness of the
interfaces in the oxide junctions \cite{barash96:4070,riedel98:6084} and
therefore may be of academic interest.

Earlier we suggested a different scenario \cite{gabovich14:1045}. Namely, we
showed that, at some relative orientations of $S_{\mathrm{BCS}}^{d}%
-I-S_{\mathrm{BCS}}^{d}$ junction electrodes, one of them can play a role of
differential detector, which enables tiny effects connected with the thermally
induced repopulation of quasiparticle levels near the FS to be observed. In
our approach, no zero-$T$ singularity of the current could arise.

A similar situation takes place for CDWSs. Although we cannot assign a
definite sign to the combined gap~$\bar{D}$ (see Eq.~(\ref{D(T_teta)}), the
corresponding unambiguously signed $\Delta$ enters the expression for the
calculation of $I_{c}$ (formulas (\ref{IcGen}) and \ref{P(T)}). In this sense,
the FS of the CDWS \textquotedblleft remembers\textquotedblright\ the specific
$\Delta$-sign at every of its points and, thus, can also serve as a
differential detector of the current at definite electrode orientations. As a
result, the dependences $I_{c}(T)$ both for symmetric $S_{\mathrm{CDW}N}%
^{d}-I-S_{\mathrm{CDW}N}^{d}$ and nonsymmetric $S_{\mathrm{CDW}N}%
^{d}-I-S_{\mathrm{BCS}}^{d}$ junctions can also by nonmonotonic and even
sign-changing functions. Unlike the $S_{\mathrm{BCS}}^{d}-I-S_{\mathrm{BCS}%
}^{d}$ junctions, for which the $I_{c}(T)$-behavior could depend only on the
orientation angles of both electrodes ($\gamma$ and $\gamma^{\prime}$), now
the other parameters responsible for the superconducting and combined
gaps---these are $\sigma_{0}$ and $\alpha$---become relevant. In Figs.~4 and
5, the $i_{c}(\tau)$ dependences are shown for various fixed $\alpha$ and
$\sigma_{0}$, respectively, both for the \textquotedblleft
checkerboard\textquotedblright\ and \textquotedblleft
unidirectional\textquotedblright\ CDW geometry. We would like to attract
attention to the fact that those dependences are rather sensitive to the
electrode orientations (see the relevant illustration in Fig.~6), so that it
might be laborious to find a suitable experimental configuration.

The key issue is that the parameters $\sigma_{0}$ and/or $\alpha$ can be
(simultaneously) varied by doping. Hence, doping CDWS electrodes and keeping
their orientations fixed, we could change even the character of the $I_{c}(T)
$ dependence: monotonic, nonmonotonic, and sign-changing. Provided the
corresponding set of parameters, we could transform the same junction, say,
from the 0-state into the $\pi$-one by varying the temperature only.


\subsection{Anomalous doping dependence of $I_{c}$}

Now, let the electrode orientations be fixed by the experimentalist
\cite{smilde05:257001,kirtley06:190} and the temperature be zero (for
simplicity), but the both parameters $\alpha$ and $\sigma_{0}$ can be varied
(by doping). In Figs.~7 and 8, the dependences of the dimensionless order
parameters $\delta(0)=\Delta(T=0)/\Delta_{0}$ and $\sigma(0)=\Sigma
(T=0)/\Delta_{0}$ on $\alpha$ and $\sigma_{0}$ are exhibited for both analyzed
CDW structures. One can see that, in every cross-section $\alpha
=\mathrm{const}$ or $\sigma_{0}=\mathrm{const}$, both $\delta(0)$ and
$\sigma(0)$ profiles are monotonic. At first glance, the Josephson tunnel
current should also demonstrate such a behavior. However, our previous
calculations \cite{gabovich13:104503,gabovich13:301,gabovich14:284} showed
that it is so when the orientations of $S_{\mathrm{CDW}N}^{d}$ electrodes in
the $S_{\mathrm{CDW}N}^{d}-I-S_{\mathrm{CDW}N}^{d}$ junction are close or
rotated by about 90$^{\circ}$ with respect to each other, i.e. when the
superconducting lobes strongly overlap in the momentum space and make
contributions of the same sign to the current. But if they are oriented in
such a way that mutually form a kind of differential detector for monitoring
the states at the gapped and non-gapped FS sections, contributions with
different signs cancel each other and more tiny effects become observable.
Such a conclusion can already be made from Figs.~4 and 5.

Really, as is illustrated by Figs.~7 and 8, in the limiting cases---$\sigma
_{0}\rightarrow\infty$ for both kinds of CDWs, and, if $\sigma_{0}\geq
\sqrt{\mathrm{e}}/2\approx0.824$ (here, $\mathrm{e}$ is the Euler constant),
$\alpha\rightarrow\pi/4$ at $N=4$ or $\pi/2$ at $N=2$ \cite{ekino11:385701}%
---we have $\delta(0)\rightarrow0$. Then, according to formulas (\ref{IcGen})
and \ref{P(T)}), $I_{c}$ also vanishes. Therefore, if the current crosses the
point $i_{c}=0$ at some values of parameters $\alpha$ or $\sigma_{0}$
different from their limiting ones, (i)~the current behavior becomes
nontrivial, because larger values of $\alpha$ and $\sigma_{0}$, which are
accompanied by smaller values of the superconducting order parameter $\delta$,
lead to the current enhancement. Nevertheless, as $\alpha$ or $\sigma_{0}$
grows further towards its corresponding limiting values, the current must
sooner or later begin to decrease by the absolute value.

This conclusion is confirmed by Figs.~9 and 10, where the dependences
$i_{c}(\sigma_{0},\alpha=\mathrm{const})$ and $i_{c}(\alpha,\sigma
_{0}=\mathrm{const})$ at $T=0$ are shown. While analyzing those figures, the
following consideration should be taken into account. Namely, we suppose that
gradual doping monotonically affects the parameters $\alpha$ and $\sigma_{0}$
of $S_{\mathrm{CDW}N}^{d}$ superconductors. Specific calculations (Figs.~9 and
10) were made assuming that only one of the control parameters, $\alpha$ or
$\sigma_{0}$, changes, which is most likely not true in the real experiment.
However, the presented results testify that each of those parameters
differently affects the current. Moreover, underdoping is usually accompanied
by the increase of both $\alpha$ and $\Sigma$ (proportional to the structural
phase transition temperature, i.e. the pseudogap appearance temperature,
$T^{\ast}$) \cite{lee07:81,gabovich10:681070,vishik12:18332,hashimoto14:483}.
Therefore, the situation when the doping-induced simultaneous changes in the
values of $\alpha$ and $\Sigma_{0}$ would lead to their mutual compensation
seems improbable. Accordingly, we believe that the proposed experiments may be
useful in one more, this time indirect, technique to probe CDWs in
high-$T_{c}$ oxides. In particular, the oscillating dependences $i_{c}%
(\alpha)$ depicted in Fig.~10b, if reproduced in the experiment, will be
certain to prove the interplay between the superconducting order parameter and
another, competing, one; here, the latter is considered theoretically to be
associated with CDWs.


\section{Conclusions}

In the two-dimensional model appropriate for cuprates, we calculated the
dependences of the stationary critical Josephson tunnel current $I_{c}$ in
junctions involving $d$-wave superconductors with CDWs on the temperature, the
CDW parameters, and the electrode orientation angles with respect to the
junction plane. It was shown that the intertwining of the CDW and
superconducting order parameters leads to peculiar dependences of $I_{c}$,
which reflect the existence of CDW gapping. The peculiarities become
especially salient when the crystal configurations on the both sides of the
sandwich make the overall current extremely sensitive to the overlap between
the superconducting lobes and the CDW sectors. In this case, the whole
structure can be considered as a differential tool suitable to detect CDWs.
Doping serves here as a control process to reveal the CDW manifestations. Such
configurations have already been created for YBa$_{2}$Cu$_{3}$O$_{7-\delta}$
\cite{smilde05:257001,kirtley06:190} and may be used to check the predictions
of our theory.


\begin{acknowledgments}

The work was partially supported by the Project N~8 of the 2012--2014
Scientific Cooperation Agreement between Poland and Ukraine. MSL was also
supported by the Narodowe Centrum Nauki in Poland (grant
No.~2011/01/B/NZ1/01622). The authors are grateful to Alexander Kasatkin
(Institute of Metal Physics, Kyiv), and Fedor Kusmartsev and Boris Chesca
(Loughborough University, Loughborough) for useful discussion.

\end{acknowledgments}


\newpage

\begin{figure}
  \centering
  \includegraphics[width=\figw]{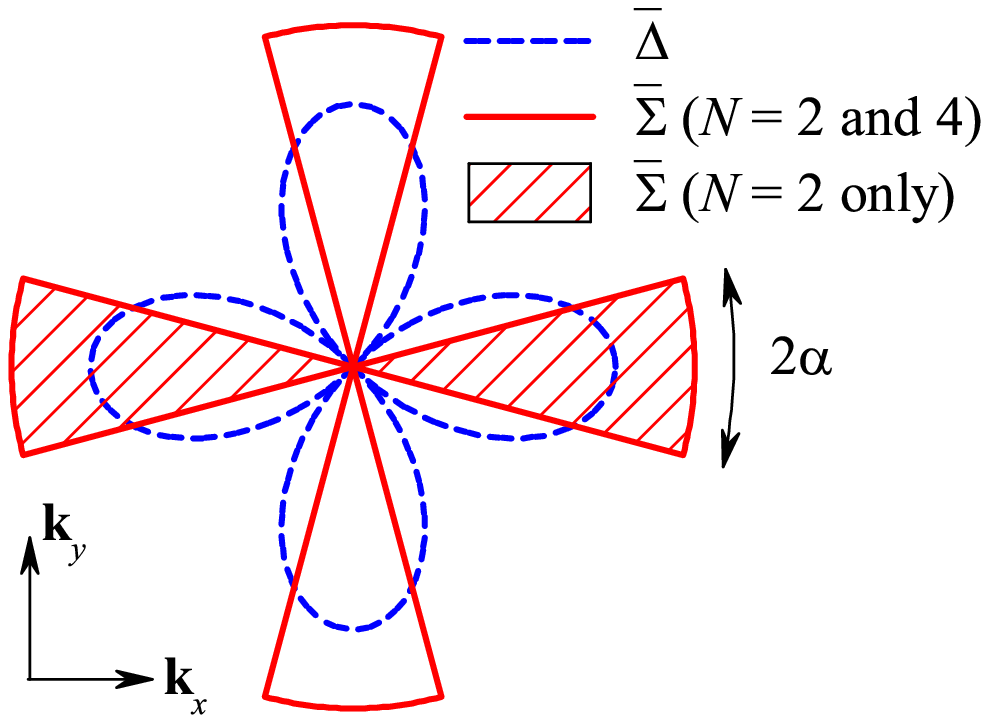}(a)
  \includegraphics[width=5cm]{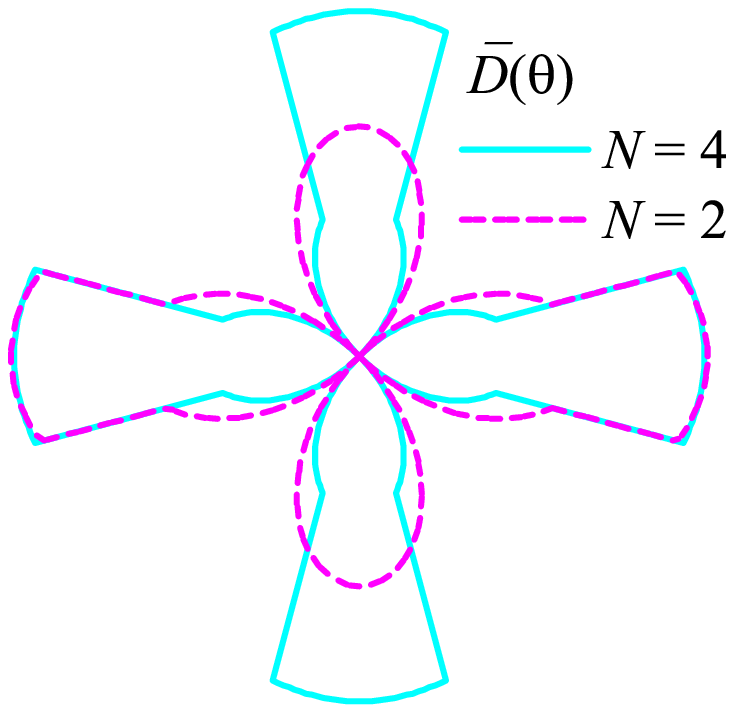}(b)  
  \\
  \caption{(a)~Superconducting, $\bar{\Delta}(\theta)$, and dielectric,
$\bar{\Sigma}(\theta)$, order parameter profiles of the partially gapped
$d$-wave charge-density-wave (CDW) superconductor. $N$ is the number of CDW
sectors with the width $2\alpha$ each. (b)~The corresponding energy-gap
contours (gap roses).}
\end{figure}

\begin{figure}
  \centering
  \includegraphics[width=10cm]{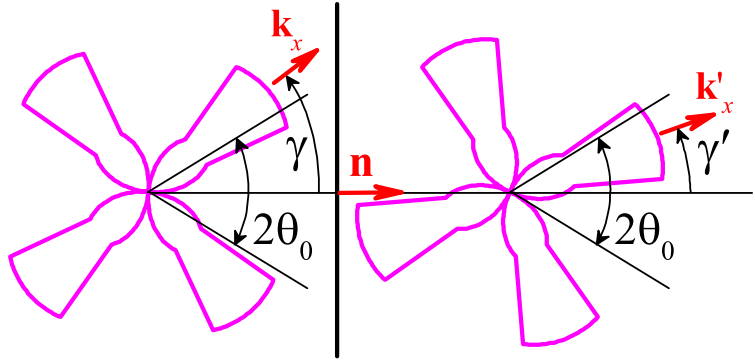}
  \\
  \caption{Configuration of symmetric Josephson junction between identical
S$_{\mathrm{CDW4}}^{d}$'s. See further explanations in the text.}
\end{figure}

\begin{figure}
  \centering
  \includegraphics[width=\figw]{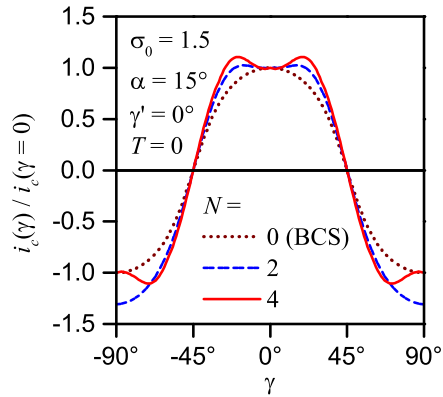}
  \\
  \caption{Orientation dependences of the reduced critical Josephson current for
the symmetric junction.}
\end{figure}

\begin{figure}
  \centering
  \includegraphics[width=\figw]{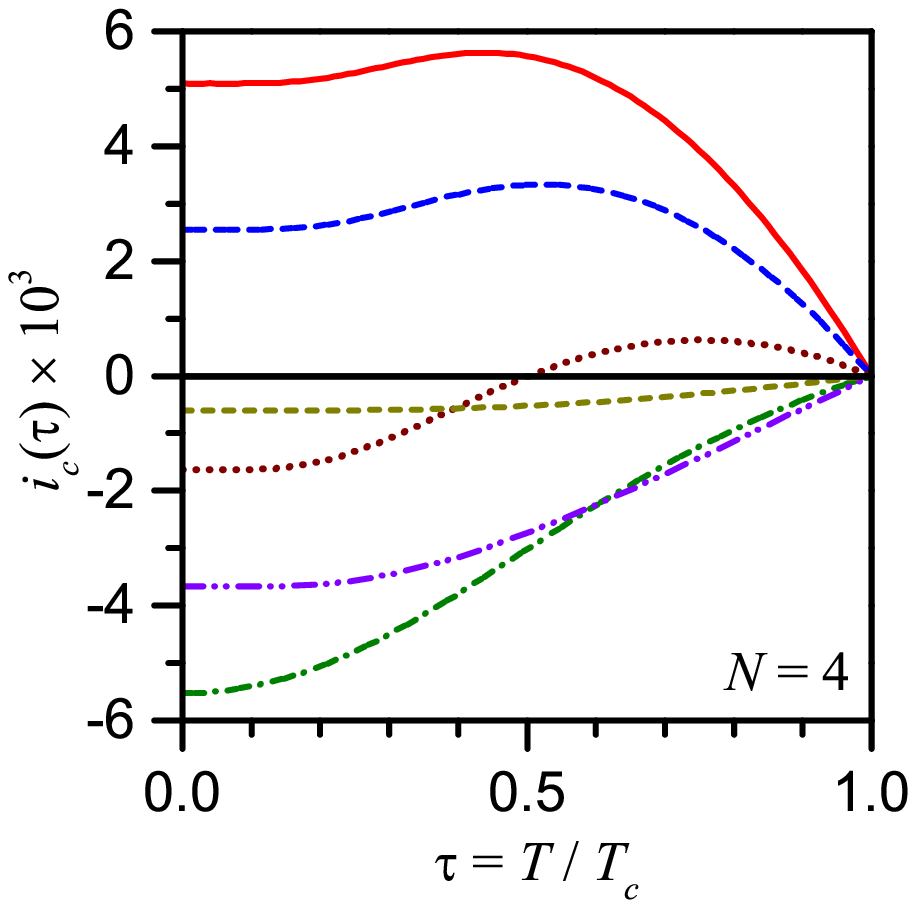}(a)
  \includegraphics[width=\figw]{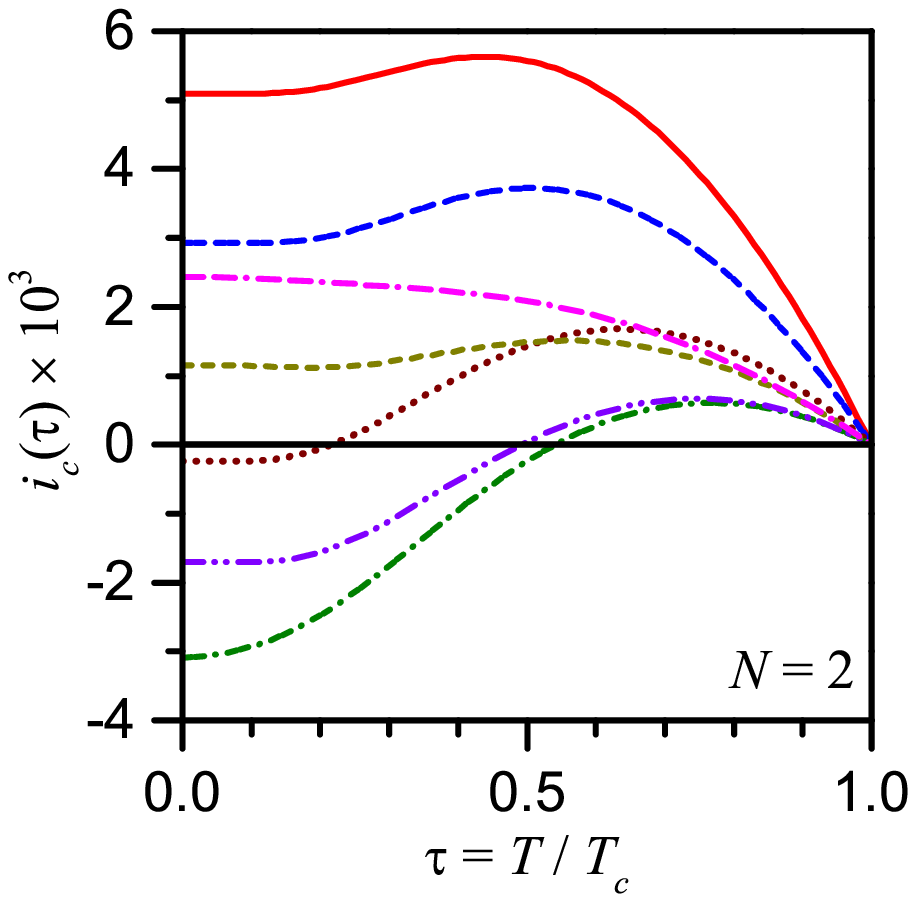}(b)
  \\
  \caption{Temperature dependences of the Josephson current for various numbers
of CDW sectors $N=4$ (a) and 2 (b), and their widths $\alpha=0$ (solid), 5
(dashed), 10 (dotted), 15 (dash-dotted), 20 (dash-dot-dotted), 25
(short-dashed), and $30^{\circ}$ (dash-dash-dotted). $\sigma_{0}=1.3$,
$\gamma=15^{\circ}$, $\gamma^{\prime}=45^{\circ}$, $\theta_{0}=10^{\circ}$.
See further explanations in the text.}
\end{figure}

\begin{figure}
  \centering
  \includegraphics[width=\figw]{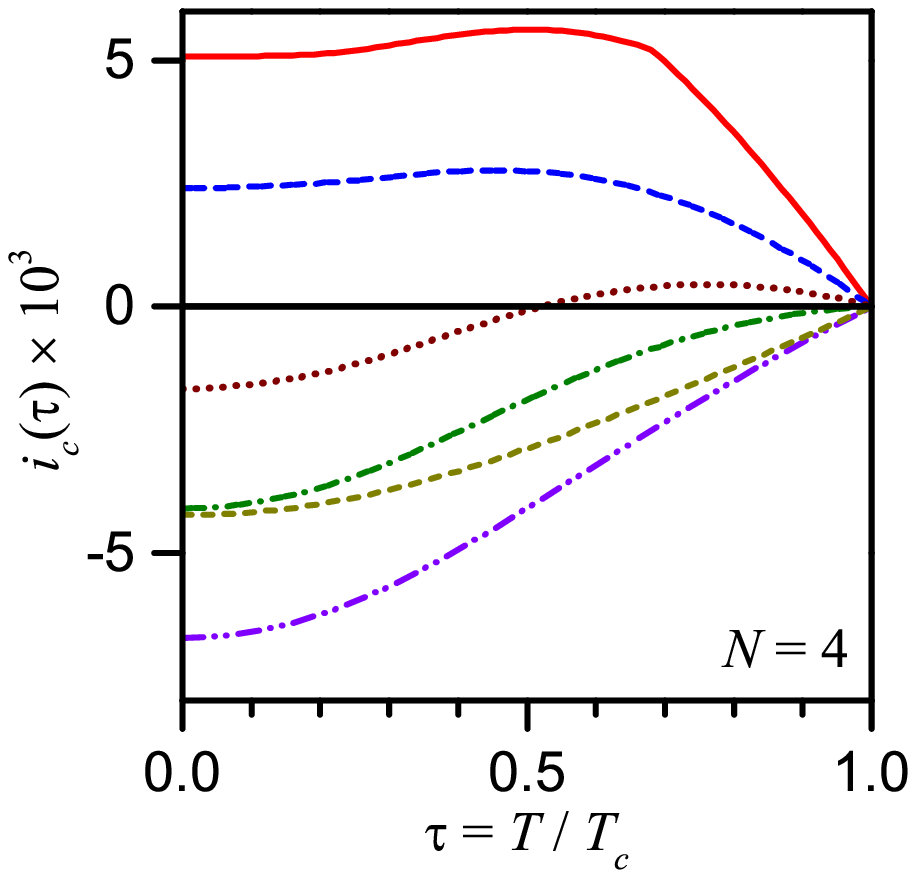}(a)
  \includegraphics[width=\figw]{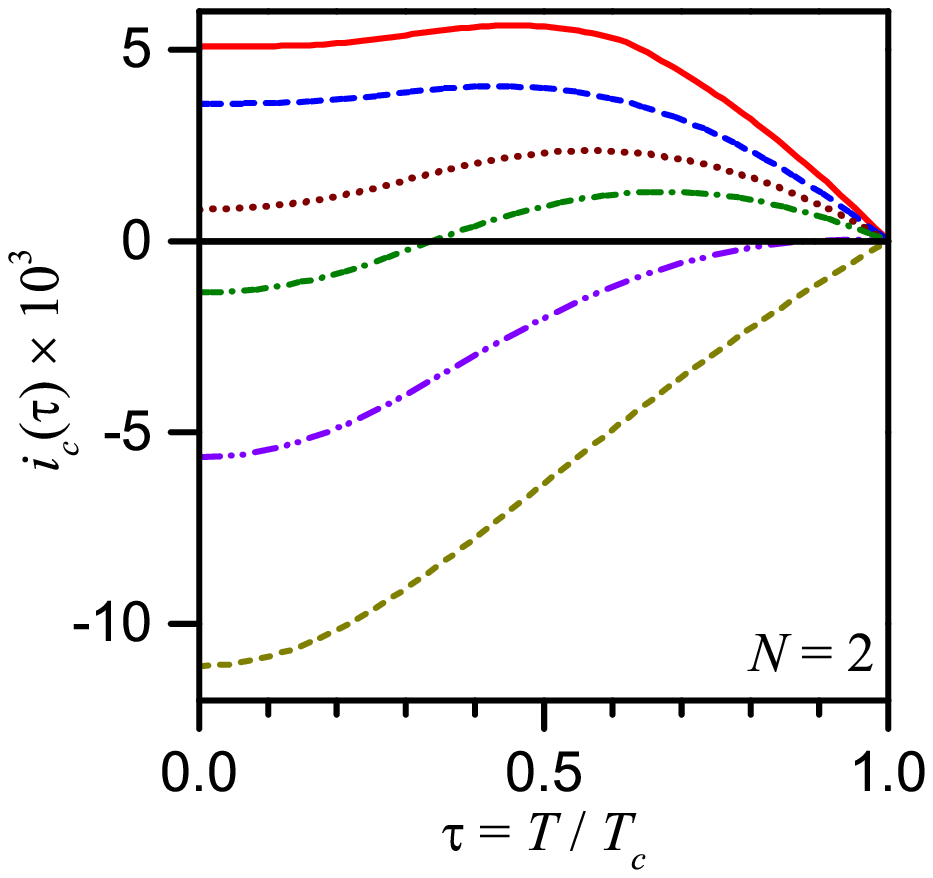}(b)
  \\
  \caption{The same as in Fig. 4, but for $\alpha=15^{\circ}$ and various
$\sigma_{0}=0.9$ (solid), 1 (dashed), 1.1 (dotted), 1.3 (dash-dotted), 1.5
(dash-dot-dotted), and 3 (short-dashed).}
\end{figure}

\begin{figure}
  \centering
  \includegraphics[width=\figw]{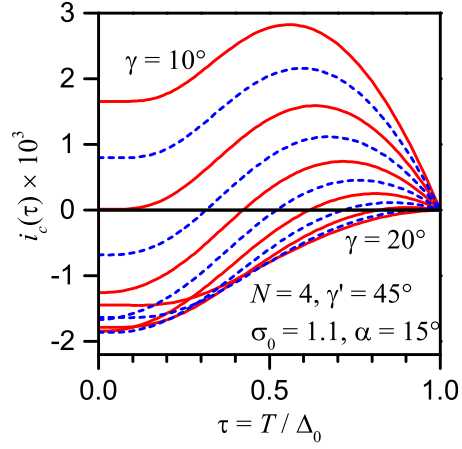}
  \\
  \caption{The same as in Fig. 4a, but for $\alpha=15^{\circ}$ and $\sigma
_{0}=1.1$ and various $10^{\circ}\leq\gamma\leq20^{\circ}$.
}
\end{figure}

\begin{figure}
  \centering
  \includegraphics[width=\figw]{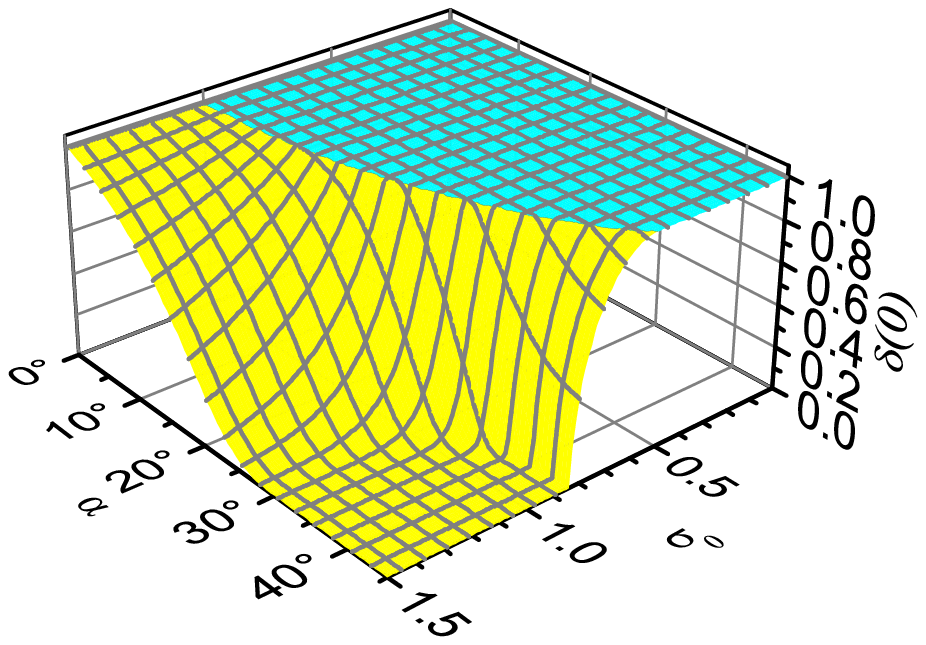}(a)
  \includegraphics[width=\figw]{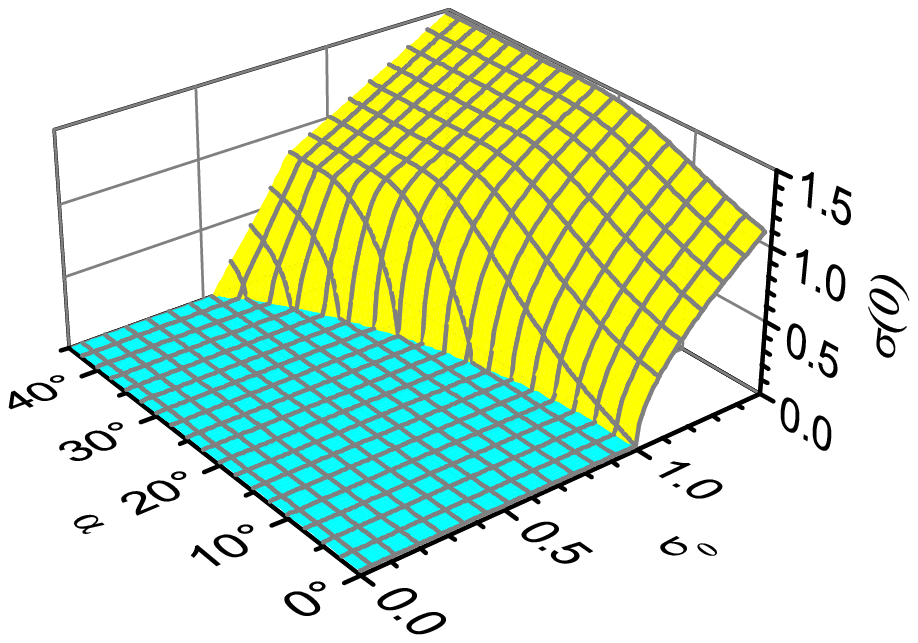}(b)
  \\
  \caption{Dependences of the normalized zero-temperature order parameters
$\delta(0)$\ (a) and\ $\sigma(0)\ $(b) for the S$_{\mathrm{CDW4}}^{d}$
superconductor on $\alpha$ and $\sigma_{0}$.}
\end{figure}

\begin{figure}
  \centering
  \includegraphics[width=\figw]{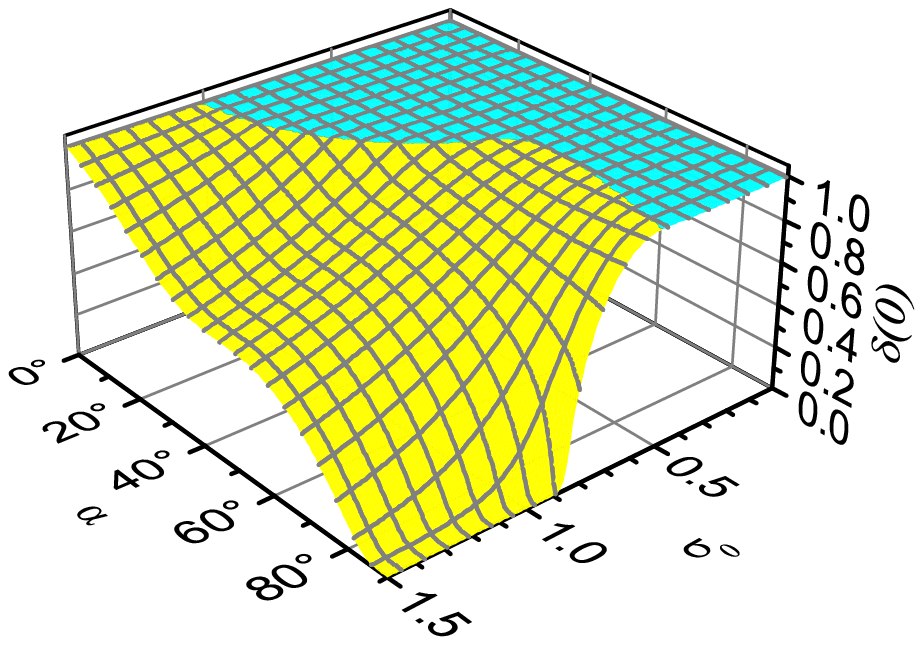}(a)
  \includegraphics[width=\figw]{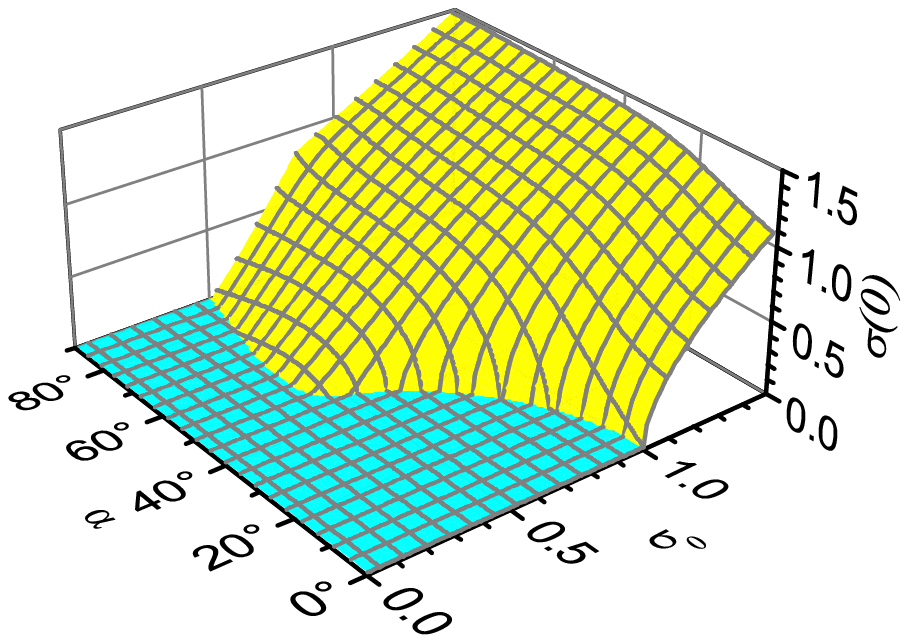}(b)
  \\
  \caption{The same as in Fig. 7, but for the S$_{\mathrm{CDW2}}^{d}$ superconductor.}
\end{figure}

\begin{figure}
  \centering
  \includegraphics[width=\figw]{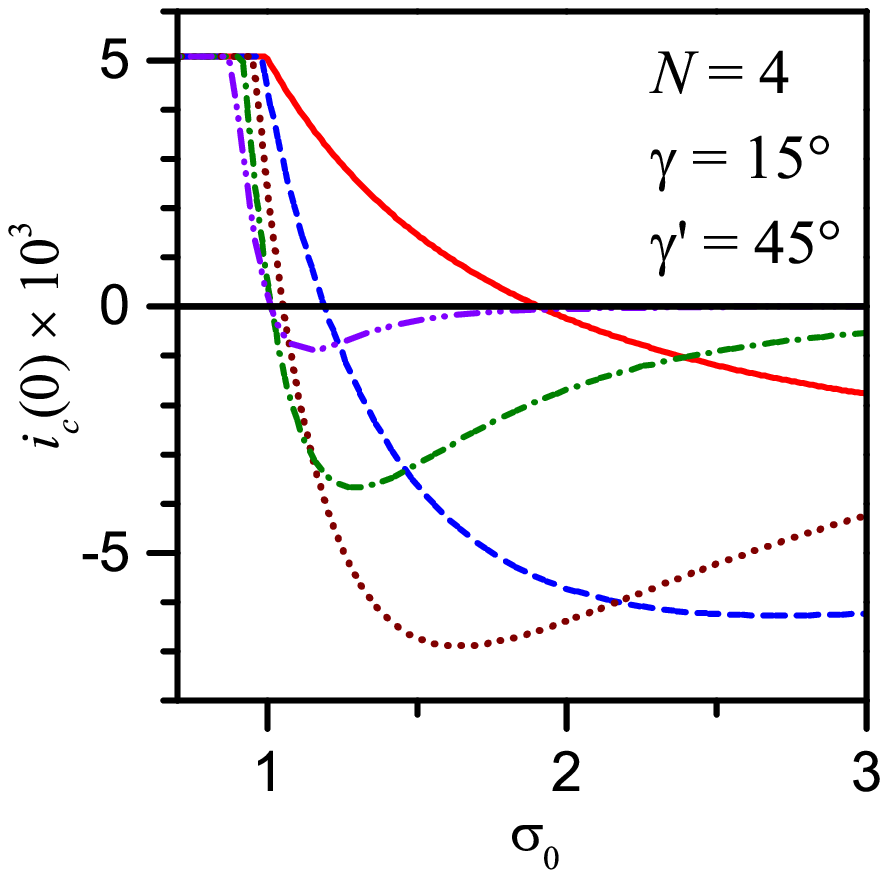}(a)
  \includegraphics[width=\figw]{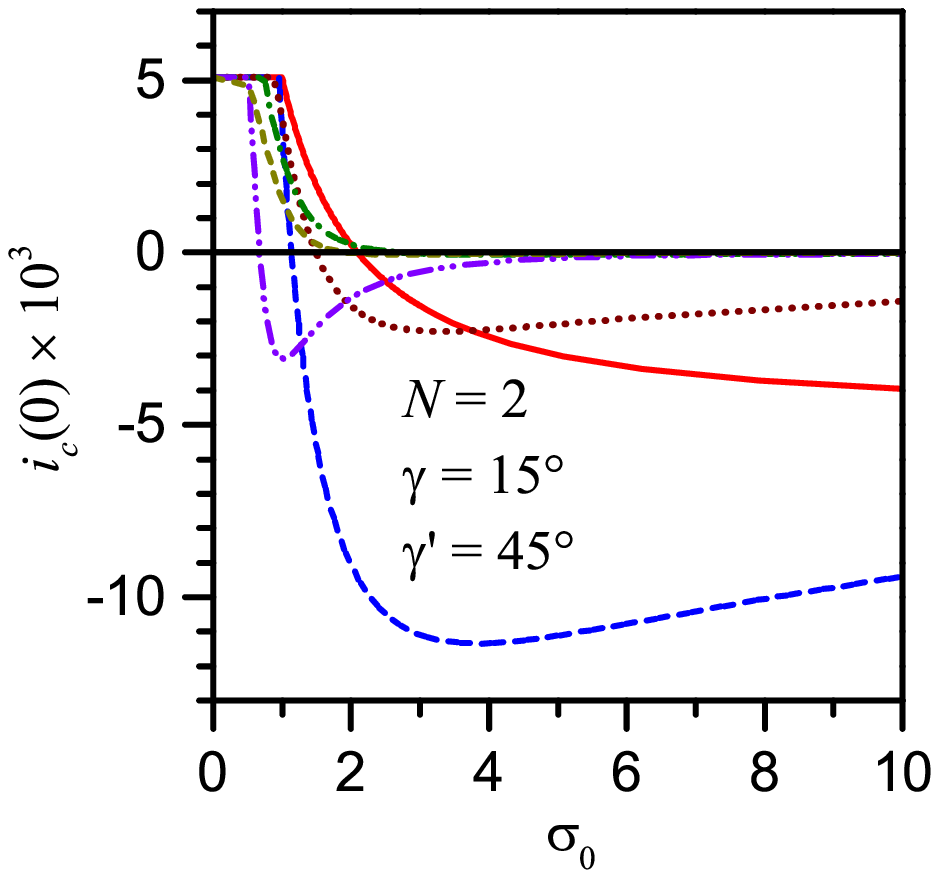}(b)
  \\
  \caption{Dependences of the normalized zero-temperature Josephson current on
$\sigma_{0}$ for $N=4$ (a) and $N=2$ (b) CDW configurations and various
$\alpha$'s:~(a) $\alpha=$ 5 (solid), 10 (dashed), 15 (dotted), 20
(dash-dotted), and 25$^{\circ}$ (dash-dot-dotted); (b)~$\alpha=$ 5 (solid), 15
(dashed), 25 (dotted), 35 (dash-dotted), 45 (dash-dot-dotted), and 55$^{\circ
}$ (short-dashed).}
\end{figure}

\begin{figure}
  \centering
  \includegraphics[width=\figw]{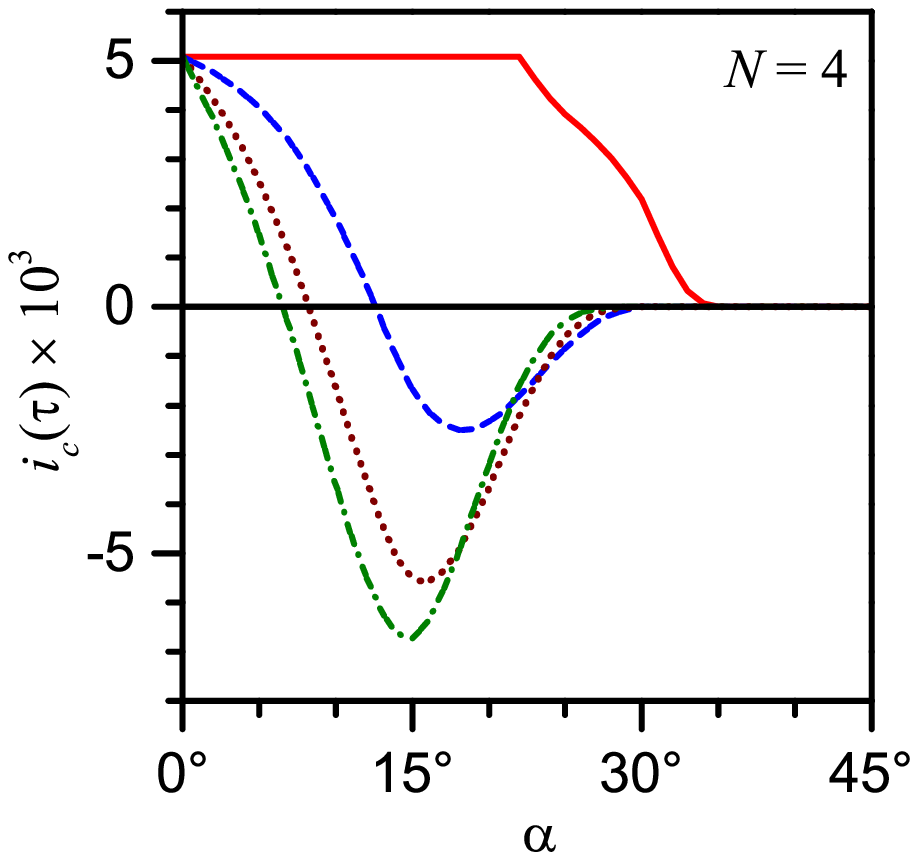}(a)
  \includegraphics[width=\figw]{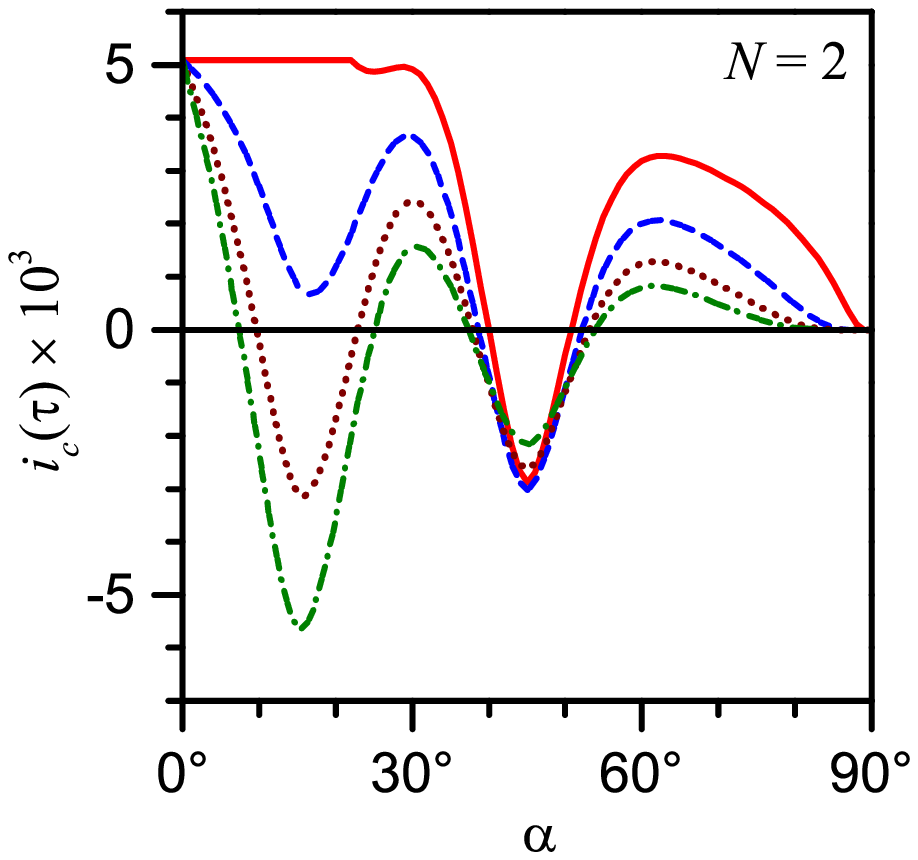}(b)
  \\
  \caption{Dependences of the normalized zero-temperature Josephson current on
$\alpha$ for $N=4$ (a) and $N=2$ (b) CDW configurations and various
$\sigma_{0}=0.9$ (solid), 1,1 (dashed), 1.3 (dotted), 1.5 (dash-dotted).
$\gamma=15^{\circ}$ and $\gamma^{\prime}=45^{\circ}$.}
\end{figure}

\end{document}